\documentclass[12pt]{article}
\input epsf.sty
\def\ZZZ{{\hbox{ Z\kern-1.6mm Z}}}

\newcommand{\bA}{{\bf A}}
\newcommand{\tx}{\wt x}

\newcommand{\ve}{\varepsilon}
\newcommand{\tl}{\wt\lambda}

\newcommand{\VV}{{\cal V}}

\newcommand{\KK}{{\cal K}}

\newcommand{\FF}{{\cal F}}
\newcommand{\HH}{{\cal H}}

\newcommand{\EE}{{\cal E}}
\newcommand{\LL}{{\cal L}}

\newcommand{\wt}{\widetilde}

\newcommand{\bd}{\bar{\rm D}}
\newcommand{\RR}{{\cal R}}

\newcommand{\TT}{{\cal T}}

\newcommand{\be}{\begin{equation}}
\newcommand{\ee}{\end{equation}}
\newcommand{\ben}{\begin{eqnarray}\displaystyle}
\newcommand{\een}{\end{eqnarray}}
\newcommand{\refb}[1]{(\ref{#1})}
\newcommand{\p}{\partial}
\newcommand{\sectiono}[1]{\section{#1}\setcounter{equation}{0}}

\def\one{{\hbox{ 1\kern-.8mm l}}}
\def\zero{{\hbox{ 0\kern-1.5mm 0}}}
 
\begin{document}
{}~
{}~
\hfill\vbox{\hbox{hep-th/0305011}
}\break
 
\vskip .6cm
\centerline{\Large \bf 
Open and Closed Strings from}

\vskip .4cm
\centerline{\Large \bf Unstable D-branes}
\vskip .6cm
\medskip

\vspace*{4.0ex}
 
\centerline{\large \rm
Ashoke Sen}
 
\vspace*{4.0ex}

\centerline{\large \it Harish-Chandra Research Institute}

\centerline{\large \it  Chhatnag Road, Jhusi,
Allahabad 211019, INDIA}
 
\centerline{E-mail: ashoke.sen@cern.ch,
sen@mri.ernet.in}
 
\vspace*{5.0ex}
 
\centerline{\bf Abstract} \bigskip

The tachyon effective field theory describing the dynamics of a
non-BPS D-$p$-brane has electric flux tube solutions where the
electric field is at its critical value and the tachyon is at its
vacuum. It has been suggested that these solutions have the
interpretation of fundamental strings. We show that in order that an
electric flux tube can `end' on a kink solution representing a BPS
D-$(p-1)$-brane, the electric flux must be embedded in a tubular
region inside which the tachyon is finite rather than at its vacuum
where it is infinite. Energetic consideration then forces the
transverse `area' of this tube to vanish.  We suggest a possible
interpretation of the original electric flux tube solutions around
the tachyon vacuum as well as of tachyon matter as system of closed
strings at density far above the Hagedorn density.

\vfill \eject
 
\baselineskip=18pt

\tableofcontents

\sectiono{Introduction} \label{s1}

Study of various aspects of tachyon dynamics on a non-BPS D-$p$-brane of 
type
IIA or IIB superstring theory has led to some understanding of the
tachyon effective 
action\cite{9909062,0003122,0003221,0004106,0204143,0209122} describing 
the dynamics of these branes.
The bosonic 
part of this effective action, describing the dynamics of the 
tachyon field on a non-BPS D$p$-brane of type IIA or IIB superstring 
theory, 
is given 
by
\ben \label{ez1}
S &=& \int d^{p+1} x \, \LL\, , \nonumber \\
\LL &=& - V(T) \, \sqrt{-\det \bA}
\, ,
\een
where
\be \label{ey2}
\bA_{\mu\nu} = \eta_{\mu\nu} + \p_\mu T \p_\nu T + 
\p_\mu Y^I \p_\nu Y^I + 
F_{\mu\nu}\, ,
\ee
\be \label{ey2a}
F_{\mu\nu} = \p_\mu A_\nu - \p_\nu A_\mu\, .
\ee
$A_\mu$ and $Y^I$ for $0\le \mu, \nu\le p$, $(p+1)\le I\le 9$ are the 
gauge and the transverse scalar fields on 
the world-volume of the non-BPS 
brane, and $T$ is the tachyon field.
$V(T)$ is the tachyon potential which is symmetric under 
$T\to -T$, has a maximum at $T=0$ where it is equal to the tension 
$\wt\TT_p$ of the non-BPS D-$p$-brane, and has its minimum at 
$T=\pm\infty$ where it vanishes. We are 
using the convention where $\eta=diag(-1, 1, \ldots 1)$ and the 
fundamental string tension has been set equal to $(2\pi)^{-1}$ 
({\it i.e.} $\alpha'=1$). 
Refs.\cite{0204143,0301076,0303035,0303139,0304045,0304108} 
suggest the 
choice:
\be \label{evt}
V(T) = {\wt \TT_p\over \cosh (T/\sqrt{2})}\, .
\ee

Since the tachyon has a negative mass$^2$ of the order of the string
scale, the very notion of an effective action, which normally refers
to the result of integrating out the heavy modes for describing the
dynamics of light modes, is somewhat unclear here. The matter is only
made worse by the fact that there are no physical states around the
tachyon vacuum, and hence the usual method of deriving an effective
action, -- by comparing the S-matrix elements computed from string
theory with those computed from the effective action, -- does not
work. Thus one might wonder in what sense \refb{ez1} describes the
tachyon effective action on a non-BPS D-$p$-brane.  This question was
addressed and partially answered in a recent paper\cite{0304045}. As
already emphasized in this paper, the usefulness of an effective
action can also be judged by comparing the classical solution of the
equations of motion derived from the effective action with the
classical solutions in open string theory, which in turn are
described by boundary conformal field theories (BCFT). In this
respect the tachyon effective action given in \refb{ez1} has had some
remarkable success. Among the string theory results reproduced by
this effective action are the following: \begin{enumerate} \item The
effective action has a one parameter family of time dependent
solutions describing the rolling of a spatially homogeneous tachyon
towards the vacuum $T=\pm\infty$. The single parameter labelling the
solution labels different initial conditions on the tachyon field
which cannot be related by time translation. In full string theory,
this one parameter family of solutions can be realized as appropriate
marginal deformations of the BCFT describing the original non-BPS
D-$p$-brane\cite{0203211,0203265}. Furthermore, for the choice of
$V(T)$ given in \refb{evt}, the time dependence of the pressure, as
calculated in BCFT, resembles the result derived from the effective
action\cite{0303139}. However, as mentioned in
\cite{0209122,0304045,0304108}, this resemblence is only at a
superficial level. This is most easily seen by examining the energy
density and pressure at the instant when the tachyon is at rest. At
this instant the field theory answers for the energy density and
pressure are equal in magnitude but differ by a sign. No such simple
relation exists for the full stringy answer. This is not necessarily
a contradiction, since during the initial stages of evolution the
second and higher derivative corrections to the action, which are not
included in \refb{ez1}, may be important. Surprisingly however, in
the limit where the tachyon begins rolling from the top of the
potential, the effective action \refb{ez1}-\refb{ey2a} with potential
\refb{evt} correctly reproduces the time evolution of the stress
tensor\cite{0303139,0304045}.

The agreement between the results derived from the effective field
theory action \refb{ez1}-\refb{evt} and the full tree level stringy
results continue to hold even in the presence of uniform background
electric field\cite{0208142,0301049,0304180}.

\item The effective action correctly gives the mass of the tachyon on
the non-BPS D-$p$-brane if we choose $V(T)$ as in \refb{evt}. In the
language of classical solutions, this can be restated by saying that
it correctly reproduces the solution of the linearized equations of
motion for the tachyon (and the massless fields) around the $T=0$
configuration.

\item The effective action has a kink solution of zero
width\cite{0011226,0012222,0208217,0301101,0301179,0303057,0304197}
representing a BPS D-$(p-1)$-brane\cite{9808141,9812031,9812135}.  
Furthermore, the world-volume action on the kink coincides with that
on a BPS D-$(p-1)$-brane\cite{0303057,0012080,0104218,0202079}. (See
also refs.\cite{0212188,0102174}.) If we choose the potential as in
\refb{ez1}, the tension of the kink also agrees with the tension of
the D-$(p-1)$-brane\cite{0303139}.

\item If we compactify one of the directions on the original
D-$p$-brane on a circle of radius $R$, then at a critical radius
$R=\sqrt 2$, the BCFT describing the non-BPS D-$p$-brane admits a
marginal deformation which smoothly interpolates between a non-BPS
D-$p$-brane and a BPS D-$(p-1)$-brane -- $\bd$-$(p-1)$-brane pair
situated at diametrically opposite points on the
circle\cite{9812031}.  It turns out that for the choice of potential
given in \refb{evt}, the effective action \refb{ez1}-\refb{ey2a}
correctly reproduces this property\cite{0303139}.  Namely, if we
compactify one of the space-like coordinates on a circle of radius
$\sqrt 2$, then the equations of motion admit a one parameter ($a$)
family of solutions, such that at one end of the parameter space
($a=0$) we have the configuration $T=0$ representing the original
non-BPS D-$p$-brane, while at the other end of the parameter space
$(a=\infty$) we have a kink-antikink pair situated at diametrically
opposite points on the circle, representing a D-$(p-1)$-brane --
$\bd$-$(p-1)$-brane pair.

\item If we consider an inhomogeneous time dependent solution by
choosing the initial condition $T=T_0\sin x$, $\dot T=0$ and let the
tachyon evolve according to the equations of motion derived from the
effective action \refb{ez1}, then the solution hits a singularity
after a {\it finite time interval} at the points $x=n\pi$ for integer
$n$\cite{0301101}. One could ask if this is also a feature of the
corresponding BCFT. Unfortunately the BCFT describing this situation
is not exactly solvable, but the corresponding problem for
D-$p$-branes in bosonic string theory is exactly solvable, and
displays precisely the feature that the energy momentum tensor blows
up at isolated values of $x$ after a finite time
interval\cite{0207105}.

\item The equations of motion derived from the effective action also
admits electric flux tube
solutions\cite{0002223,0009061,0010240,9901159,0209034,LIND}.  These
solutions are characterized by the fact that the tachyon is at its
vacuum ($T=\infty$), the electric field $\vec E$ takes its limiting
value $|\vec E|=1$, and the electric flux is non-zero in some region
of space.  The spatially homogeneous version of these solutions can
be realized as appropriate deformations of the BCFT describing the
non-BPS D-$p$-brane with background electric field\cite{0208142,0301049}.
\label{p5}

\end{enumerate}

The localized electric flux tube solutions mentioned in item \ref{p5}
above have been proposed as candidates for describing fundamental
strings. Indeed, these solutions have many properties in common with
the fundamental string, including the quantum numbers and
dynamics\cite{0002223,0009061,0010240}.  However these solutions also
suffer from the difficulty that the flux can spread out in the
transverse directions instead of being confined into a narrow 
tube\cite{0002223,0009061,0303133}.
This will correspond to a new degree of freedom corresponding to
fattening of the fundamental string, and is contrary to the known
property of the fundamental string.

In this note we show that if an electric flux tube has one of its
ends `attached' to a kink, then as we move a distance $x$ away from
the kink along the flux tube, the tachyon cannot increase faster than
$x^{1/2}$. This result, in turn, can be used to argue that the usual
exchange interaction, expected of a fundamental string, is suppressed
for these flux tubes.  Hence we have another reason for not using
these solutions to describe a fundamental string. In order to
overcome this problem, we propose another class of solutions as
candidates for fundamental strings based on the result of
\cite{9708147,9709027,9709014,9711094}. These solutions have the
right quantum numbers and dynamics as a fundamental string, have the
ability to end on a kink, and can also have the usual exchange
interaction. The construction of these solutions requires creating a
core where the tachyon solution is away from its vacuum value, and
the electric flux is embedded in this core. Energetic consideration
then forces this core to have zero `area' in the hyper-plane
transverse to the flux.  Consequently, the electric flux is also
confined to a region of zero volume. This property is consistent with
that of the fundamental string, but this is only a partial success,
as confinement of the flux within a region of zero volume does not
necessarily imply confinement into a one dimensional subspace as
would be required if it has to describe a fundamental string. We
suggest a possible resolution of this puzzle based on higher
derivative corrections on the D-brane world volume.

In this context we also recall that in several recent papers
precisely this type of configurations involving electric flux have
been considered from another viewpoint\cite{0303133,0211090,0303172}.  
In these papers the authors studied the process of inhomogeneous
tachyon condensation on an unstable D-brane in the presence of an
electric field and argued that at the end of the condensation process
the electric flux gets confined into regions inside which the tachyon
is finite, rather than being spread out into the fat flux tube
solutions described in \cite{0009061} for which the tachyon is at its
vacuum everywhere. Based on this analysis the authors argued that
fundamental string solutions must be described by the former type of
solutions. Although we arrive at these solutions from a different
point of view, our final conclusion agrees with that of
refs.\cite{0303133,0211090,0303172}.\footnote{We should note however
that the explicit string theoretic analysis based on boundary state
has been carried out only in the context of bosonic string
theory\cite{0303133}, since a perturbation describing inhomogeneous 
rolling tachyon gives rise to a solvable BCFT only in this 
case\cite{0207105}. Thus these results are not directly applicable in the 
present context where we focus our attention on the superstring 
theory.}

These results still leave open the question: what is the physical
interpretation of the original electric flux tube solutions of
\cite{0002223,0009061,0010240,9901159,0209034,LIND} inside which the
tachyon is at its vacuum everywhere? A similar question can be asked
about the tachyon matter solution of
\cite{0203211,0203265,0204143,0301101,0301179, 0207105,0202210,
0209090,0301038,0302146,0208196,0212248,0208019,0207107,
0304163,0205085,0205098,0206102,0206212,0301137,0210221,0209222}. We
suggest a possible interpretation of these solutions as a system of
high density closed string states. Our arguments rely on the results
of \cite{0303139,0304192} where the authors argued that during the
decay of a D-brane all the energy of the brane is converted into
closed strings.  Naively one might expect that this invalidates the
open string analysis of \cite{0203211,0203265}. However, since closed
strings are automatically included in quantum open string theory, one
could argue that the effect of emission of closed strings should be
included in quantum open string theory, and need not have to be
included as a separate effect.  Thus the closed string description of
the D-brane decay should be equivalent to the quantum open string
description rather than being a replacement of the latter. The
results in quantum open string theory, in turn, should reduce to
those in the classical open string theory in the weak string coupling
limit.  For consistency, this requires that the properties of the
system of closed strings produced in the decay of a D-brane must
agree with the results of classical open string theory in the weak
string coupling limit. We show that this is indeed the case. This
suggests that in the weak coupling limit the classical tachyon matter
described in \cite{0203211,0203265} gives a description of a system
of closed strings at density of order $g^{-1}$, {\it i.e.} far above
the Hagedorn density.  Generalizing this, we also propose that the
electric flux tube solutions with electric field directed along a
compact direction describes a system of closed strings at high energy
density and high density of fundamental string winding charge. This
indicates that in general the solutions in the classical open string
theory (or tachyon effective field theory) around the tachyon vacuum
where $T$ is large everywhere give effective description of closed
strings at density of order $g^{-1}$. Interpretation of tachyon
matter in a somewhat similar spirit has been discussed independently
in \cite{0304192}.

\sectiono{Review of Kink and Flux Tube Solutions} \label{s2}

We begin by reviewing the
flux tube solutions of refs.\cite{0002223,0009061,0010240,9901159}. 
These 
are 
most easily seen in the Hamiltonian formalism given in
\cite{0009061}. We denote by $\Pi^i$ the momenta conjugate to the gauge
field components $A_i$ ($1\le i\le p$), by $P_I$ the 
momenta conjugate to
the scalar field $Y^I$, and by $\Pi^T$ the momentum conjugate to the 
tachyon field $T$:
\ben \label{ewmom}
\Pi^i &=& {\delta S \over \delta (\p_0 A_i)} = V(T)\, (\bA^{-1})^{i0}_A 
\sqrt{-\det \bA} \, , \nonumber \\
P_I &=& {\delta S \over 
\delta (\p_0 Y^I)} = V(T)\, (\bA^{-1})^{0\nu}_S \p_\nu Y^I \sqrt{-\det
\bA} \, , \nonumber \\
\Pi_T &=& {\delta S \over \delta (\p_0 T)} = V(T)\, (\bA^{-1})^{0\nu}_S 
\p_\nu T \sqrt{-\det
\bA} \, ,
\een
where the subscripts $S$ and $A$ denote the symmetric and anti-symmetric
components of a matrix respectively.
Then the Hamiltonian is given by:
\ben \label{ewidth1}
H
&=& \int d^p x \, \HH(x)\, , \qquad \HH(x) = \sqrt{\KK(x)} \nonumber 
\\
\KK(x) &=& \Pi^i \Pi^i + P_I P_I + \Pi_T \Pi_T + (\Pi^i \p_i Y^I) (\Pi^j
\p_j Y^I) + (\Pi^i \p_i T) (\Pi^j
\p_j T)\nonumber \\
&& + (F_{ij} \Pi^j + \p_i 
Y^I P_I + \p_i T \, \Pi_T) (F_{ik} \Pi^k + \p_i
Y^J P_J + \p_i T \, \Pi_T) \nonumber \\
&& + V^2 \det (h) \nonumber \\
h_{ij} &=& \delta_{ij} + F_{ij} + \p_i Y^I \p_j Y^I + \p_i T \p_j T\, .
\een
The equations of motion derived from this Hamiltonian needs to be 
supplemented by the Gauss' law constraint:
\be \label{ewidth2}
\p_i \Pi^i = 0\, .
\ee

Around the minimum of the tachyon potential at $T=\infty$ the theory 
contains solutions describing electric flux tubes. For example, the 
configuration
\be \label{ewidth2a}
\Pi^1(x) = f(x^2, \ldots x^p), \qquad A_1(x) = x^0, \qquad 
T=\infty\, ,
\ee
with all other fields and their conjugate momenta set to zero, can be 
shown to be a solution of the equations of motion, and describes 
electric flux along the $x^1$ direction. $f(x^2,\ldots x^p)$ is an 
arbitrary positive semi-definite function of the coordinates 
transverse to the direction of the 
flux. 
The total flux or fundamental string charge associated with 
this configuration is given by
\be \label{efl0}
\FF = \int dx^2 \ldots d x^p \, f(x^2, \ldots x^p)\, .
\ee
This is quantized in units of fundamental string charge.
Taking $f$ to be a delta function in these transverse coordinates 
gives a string-like configuration whose dynamics agrees with that of the 
fundamental string\cite{0009061,0010240,LIND}, but clearly the 
freedom of 
choosing an arbitrary $f$ 
shows that the flux can spread out, unlike that of a fundamental string 
which has zero width. 

As already stated,
the equations of motion derived from the Hamiltonian \refb{ewidth1} also 
has a tachyon kink 
solution\cite{0011226,0012222,0208217,0301101,0301179,0303057}. 
It has the property that it has zero 
width, and interpolates between $T=-\infty$ for $x<0$ to $T=\infty$ for 
$x>0$. For describing electric flux 
in the presence of a kink, it is useful to take the kink as the 
$a\to\infty$ limit of 
the configuration\cite{0208217}:
\be \label{ewidth2b}
T(x^p) = a F(x^p)
\ee
where $F(x)$ satisfies
\ben \label{ewidth2bb}
&&F(-x) = - F(x), \qquad F'(x) > 0 \quad \hbox{for} \quad -b < x < b,
\quad F'(\pm b) = 0, \nonumber \\
&& F(x) = F(b) \quad \hbox{for} \quad x \ge b, 
\quad  
F(x) = F(-b) \quad \hbox{for} \quad x \le -b\, ,
\een
$b$ being a constant which can be taken to be as small as we like {\it 
after 
we have taken the $a\to\infty$ limit.}
This solution has the property that $\p_i 
T$ goes to zero outside the range $-b< x^p< b$, but in the range 
$-b<x^p<b$ 
$|\p_{x^p} T|$ is infinite in the $a\to\infty$ limit, and $T(x^p)$ 
interpolates 
between $T(x^p)=-\infty$ for $x^p<0$ to $T(x^p)=\infty$ for $x^p>0$. 

This description is slightly different from the one used in 
\cite{0303057}, but
all the properties of the solution discussed in \cite{0303057} remain 
valid with this new description.
Let $\xi\equiv (\xi^0,\ldots \xi^{p-1}) = (x^0, \ldots x^{p-1})$ 
denote the world-volume coordinates on the kink.
As was shown in \cite{0303057}, the world-volume theory on the kink 
is described precisely by the Dirac-Born-Infeld (DBI) action 
involving 
massless scalar fields $y^i(\xi)$ ($p\le i\le 9$) and gauge fields 
$a_\alpha(\xi)$ ($0\le\alpha\le (p-1)$) under the identification:
\ben \label{eprofile}
&& A_p(x^p, \xi) = \phi(x^p,\xi), \qquad A_\alpha(x^p,\xi) = 
a_\alpha(\xi) - \phi(x^p,\xi) \, {\p y^p(\xi)\over \p \xi^\alpha}, 
\qquad 
\nonumber \\
&&  T(x^p,\xi) = a F(x^p-y^p(\xi))\, , \qquad
Y^I(x^p,\xi) = y^I(\xi), \quad \hbox{for} \quad (p+1)\le I\le 9 
\, ,
\een
where $\phi(x^p, \xi)$ is an arbitrary smooth function. In fact 
$A_\mu(x^p,\xi)$ ($Y^I(x^p,\xi)$) can be taken to be any smooth 
vector (scalar) field 
whose pull-back along the kink world-volume $x^p=y^p(\xi)$ is equal 
to $a_\alpha(\xi)$ ($y^I(\xi)$). Using these relations one can show 
that \ben 
\label{ewidth2c}
\Pi^s(x^p, \xi) &=& (\TT_{p-1})^{-1}\, aF'(x^p-y^p(\xi)) 
V(aF(x^p-y^p(\xi))) 
\, 
\pi^s(\xi) \nonumber \\
&=& (\TT_{p-1})^{-1}\, \pi^s(\xi) \, {\p\over \p x^p} 
U(T(x^p-y^p(\xi)))\, , 
\qquad 1\le s\le (p-1) \, ,\nonumber \\
U(T) &\equiv& \int_{-\infty}^T V(y) dy\, , \nonumber \\
\Pi^p(x^p, \xi) &=& \Pi^s(x^p,\xi) \, {\p y^p(\xi)\over \p 
\xi^s} = 
- (\TT_{p-1})^{-1}\, \pi^s(\xi) \, {\p\over \p\xi^s} \, U(T(x^p - 
y^p(\xi))\, .
\een
where
$\pi^s=(\delta S / \delta (\p_0 a_s))$
denotes the momenta conjugate to the gauge fields on the 
kink world-volume and
\be \label{ewidth2d}
\TT_{p-1} = \int_{-\infty}^\infty V(T) dT = U(\infty)\, ,
\ee
is the tension of the kink.
Since $V(T)$ falls off exponentially for large $T$, we see that the 
most of the contribution to $\Pi^s$ comes from the region where 
$F(x^p-y^p(\xi))$ is of order $1/a$, {\it i.e.} from regions of width 
$\sim 
{1\over 
a}$.

Incidentally, \refb{ewidth2c} can also be derived from an energy 
minimization principle. Consider, for example, a flat BPS 
D-$(p-1)$-brane at 
$x^p=0$, with a uniform electric flux $\pi^1$ along the $x^1$ direction. 
In the tachyon effective field theory on a non-BPS D$p$-brane, we 
could try to 
represent this by  the tachyon background given in 
\refb{ewidth2b}, with an $x^p$ dependent electric flux $\Pi^1(x^p)$ along 
$x^1$ direction. $\pi^1$ is related to $\Pi^1(x^p)$ as
\be \label{eadd1}
\pi^1 = \int d x^p \, \Pi^1 (x^p) \, .
\ee
{}From \refb{ewidth1} we now see that in the $a\to\infty$ limit the 
energy density associated with 
this configuration is given by:
\be \label{eadd2}
\EE = \int d x^p \sqrt{\left\{\Pi^1(x^p)\right\}^2 + 
\left\{V(aF(x^p)) a 
F'(x^p)\right\}^2}\, .
\ee
We want to find what $\Pi^1(x^p)$, subject to the constraint \refb{eadd1}, 
minimizes this energy density. For this we take the ansatz:
\be \label{eadd2a}
\Pi^1(x^p) = G'(x^p)\, ,
\ee
where $G$ is some function to be determined. \refb{eadd1} now gives:
\be \label{eadd3}
G(\infty) - G(-\infty) = \pi^1\, .
\ee
On the other hand, \refb{eadd2} gives
\be \label{eadd4}
\EE = \int d x^p \sqrt{ \{G'(x^p)\}^2 + \left\{V(aF(x^p)) a
F'(x^p)\right\}^2}\, .
\ee
We now minimize \refb{eadd4} with respect to $G(x^p)$ keepings the 
boundary values of $G(x^p)$ fixed. This gives:
\be \label{eadd5}
{\p\over \p x^p} \left( {G'(x^p) \over \sqrt{ \{G'(x^p)\}^2 + 
\left\{V(aF(x^p)) a
F'(x^p)\right\}^2} } \right) = 0\, .
\ee
Thus:
\be \label{eadd6}
G'(x^p) = C V(aF(x^p)) a
F'(x^p) \, ,
\ee
where $C$ is a constant. Integrating both sides over $x^p$, and using 
\refb{ewidth2d}, \refb{eadd3}, we get:
\be \label{eadd7}
\pi^1 = C \TT_{p-1}\, .
\ee
\refb{eadd2a}, \refb{eadd6}, \refb{eadd7} now give:
\be \label{eadd8}
\Pi^1 (x^p) = (\TT_{p-1})^{-1}\, aF'(x^p)V(aF(x^p)) \, \pi^1\, .
\ee
\refb{eadd8} is a special case ($y^p(\xi)=0$) of the more general 
result given in 
\refb{ewidth2c}.

\sectiono{A No Go Theorem} \label{s3}

Since a kink solution in this theory is expected to describe a 
D-$(p-1)$-brane, we would expect that fundamental strings should be able 
to end on this D-$(p-1)$-brane.
Since the end of the fundamental string carries electric 
charge under the $U(1)$ gauge field living on the D-$(p-1)$-brane, this 
will give rise to a gauge field background on the D-$(p-1)$-brane. 
According to \refb{ewidth2c} this implies that the electric flux 
associated with the fundamental string should be able to penetrate to the 
core of the kink where $F(x^p-y^p(\xi))\sim a^{-1}$, and $T=a 
F(x^p-y^p(\xi))$ is finite.
We shall now show that there are strong constraints on finding 
electric flux tube solutions of this type. 

We begin by noting that
$\KK(x)$ given in eq.\refb{ewidth1} is a sum over a set of terms each of 
which is positive 
semi-definite. This allows us to put a lower bound to the total energy 
associated with any configuration as follows:
\be \label{ewidth3}
E = H \ge \int d^p x \, \sqrt{(\Pi^i \p_i T) (\Pi^j
\p_j T)} = \int d^p x \, |\Pi^i \p_i T| = \int d^p x \, |\p_i (\Pi^i 
T)|\, .
\ee
In the last step we have used the Gauss law constraint \refb{ewidth2}.
Let us now evaluate the contribution to the right hand side of 
\refb{ewidth3} from a narrow tube 
around an electric flux line, with the walls 
of the tube being parallel to 
the flux line, and the two ends $A$ and $B$ capped by disks of 
cross-section $d\sigma_A$ and $d\sigma_B$ orthogonal to the flux lines. 
Let $T_A$ and $T_B$ be the values of the tachyon field at 
the two ends and 
$\Pi_A$ and $\Pi_B$ be the magnitudes of $|\vec\Pi|$ at the two ends. 
Then, since there is no leakage of flux 
through the wall of the tube, the 
Gauss law constraint \refb{ewidth2} gives:
\be \label{efl1}
\Pi_A \, d\sigma_A = \Pi_B \, d\sigma_B \equiv d\FF\, ,
\ee
where $d\FF$ denotes the total flux flowing along the tube. On the other 
hand, evaluating the contribution to the right hand side of 
\refb{ewidth3} from this tube we see that the 
total energy $dE$ contained in 
this 
tube has a lower bound of the form:
\be \label{efl2}
dE \ge |\Pi_A \, T_A \, d\sigma_A - \Pi_B \, T_B \, d\sigma_B| = |d\FF 
(T_A - T_B)|\, .
\ee
This shows that if for a finite amount of flux flowing along a tube the 
value of the tachyon 
along a flux line changes by an infinite amount, -- as will be the 
case 
if the electric flux enters the core of the kink from outside where 
$T=\infty$, -- then it costs an 
infinite amount of energy.
Put another way, the electric flux tubes of the kind described in 
\refb{ewidth2a} are repelled by the tachyon kink for $|x|<b$, since 
$|\vec\nabla T|$ blows up in this range.
Hence such flux 
tubes cannot `end' on a D-brane.

Note, however, that this argument only prevents the tachyon from 
changing from a finite value to $\infty$ along a flux line {\it 
within a finite distance.} An infinite length flux tube can have at 
one end a finite $T$ and at the other end infinite $T$. To see how 
this is possible, let us take a flux line along (say) the $x$ 
direction, carrying total electric flux $\Pi$, and let $T(x)$ denote 
the tachyon profile along $x$. 
We shall take the tachyon to be large so that the term involving the 
tachyon potential can be ignored.
Then 
the total integrated energy associated with the flux line is given 
by:
\be \label{eint}
\int dx |\Pi| \sqrt{1 + (\p_x T)^2} \le \int dx |\Pi(x)| (1 + 
{1\over 2} (\p_x T)^2 )\, .
\ee
In the right hand side of \refb{eint} the first term $\int dx\, 
|\Pi(x)|$ is just the 
energy cost due to the fundamental string tension which is always 
present. Thus the excess contribution to the energy due to the 
variation of $T$ 
along the flux line is bounded from above by
\be \label{eint1}
{1\over 2} \, \int \,  dx \, |\Pi|\, (\p_x T)^2 \, .
\ee
Now suppose $T=T_0$ at $x=0$ where $T_0$ is some arbitrary large but  
finite 
constant. If we take {\it e.g.} $T$ to vary along the flux tube as:
\be \label{eint2}
T(x) = T_0 - \alpha + \alpha \, (1+x)^\beta
\ee
for some constants $\alpha$ and $\beta$ with $0<\beta<{1\over 2}$, 
then $T$ approaches $\infty$ as $x\to\infty$. On the other hand 
\refb{eint1} shows that the 
total energy cost for this configuration is bounded from above by:
\be \label{eint3}
{1\over 2} \, |\Pi|\, {\alpha^2 \beta^2\over  1 - 2\beta} \, .
\ee
This can be made as small as we like by taking $\alpha$ sufficiently 
small.
Thus we see that at little cost in energy, it is possible to have 
configurations where the tachyon grows slowly towards infinity as we 
move along the flux line away from the plane of the kink. We can put 
a bound on how fast the tachyon can grow by requiring that the excess 
energy
\be \label{ecomp0}
T^{excess}_{00} \equiv \int^\infty dx |\Pi| (\sqrt{1 + (\p_x T)^2} - 
1) \, ,
\ee
be finite. If $T \sim x^\beta$ for large $x$, then this gives a bound 
$\beta< {1\over 2}$, and the excess energy density behaves for large 
$x$ 
as:
\be \label{ecomp1}
T^{excess}_{00} \sim x^{2\beta -2}\, .
\ee

The results of this section hold also for a configuration 
describing a flux tube passing through the kink rather than ending on 
it, since in order to pass through the kink the flux must travel 
through a region inside which the tachyon is finite. Since there are 
explicit classical solutions describing electric flux passing through 
a kink, both as exact classical solutions in open string 
theory\cite{0208142,0301049}, and as classical solutions in the tachyon 
effective field theory\cite{0304180}, we shall examine these 
solutions in the next section and explicitly verify the various 
bounds derived in this section.

\sectiono{Boundary Conformal Field Theory Analysis} \label{s3a}

We shall now 
construct, using BCFT techniques, a periodic array of kink-antikink 
pairs on a 
non-BPS D$p$-brane in the 
presence of an electric field in direction transverse to the 
kink world-volume, and show that the results are consistent with the 
analysis of section \ref{s3}.
We begin by reviewing the case without the electric field. This
construction requires us to switch 
on a 
tachyon field configuration of the form
$T \propto \tl \cos(x /\sqrt 2)$\cite{9808141,9812031}, where $x$ 
denotes a particular direction on the D-$p$-brane. For definiteness 
we shall take $x=x^p$. The 
space-time energy-momentum 
tensor associated with this deformed BCFT can be obtained by 
examining the associated boundary state, and can in fact be read out 
using a Wick rotation of the results in \cite{0203265}:
\be \label{ebc1}
T_{00} = \wt\TT_p f(x), \quad T_{pp} = -{1\over 2} \wt\TT_p \left(1 
+ 
\cos(2\pi\tl)\right)\, , \quad T_{ij} = - \wt\TT_p f(x) \delta_{ij} 
\quad 
\hbox{for} 
\quad 1\le i,j\le (p-1)\, ,
\ee
where
\ben \label{ebc2}
f(x) &=& {1\over 1 + \sin^2(\pi\tl) e^{i\sqrt 2 x} } + {1\over 1 + 
\sin^2(\pi\tl) e^{-i\sqrt 2 x} } - 1 \nonumber \\
&=& {\left( 1 + \sin^2(\pi\tl)\right) \cos^2(\pi\tl)\over 
\cos^4(\pi\tl) + 2 
\sin^2(\pi\tl) \left(1 + \cos(\sqrt 2 x)\right)}
\een
{}From this we see that as $\tl\to 1/2$, $f(x)$ vanishes everywhere 
except in the neighbourhood of $x=(2n+1) \pi/\sqrt 2$ for integer 
$n$. A close examination shows 
that in the $\tl\to 1/2$ limit $T_{00}$ receives a delta-function 
contribution equal to
\be \label{ebc3}
T_{00} = \TT_{p-1} \, \sum_n \, \delta (x^p - (2n+1) \pi/\sqrt 2)\, ,
\ee
where $\TT_{p-1} = \sqrt 2 \, \pi \, \wt \TT_p$ is the tension of a 
BPS D-$(p-1)$-brane. Thus at $\tl =1/2$ the BCFT describes an array 
of
D-$(p-1)$-brane $\bd$-$(p-1)$-brane 
pairs\cite{9812031}.\footnote{While the 
energy-momentum tensor 
does not distinguish a D-$(p-1)$-brane from a $\bd$-$(p-1)$-brane, 
they can be distinguished by examining the expression for the RR 
charge density.}
If we take $\tl\simeq 1/2$ instead of $\tl=1/2$, it 
is easy to see 
from \refb{ebc2} that the energy density is concentrated in a region 
of width 
\be \label{ebc4}
\Delta \sim \cos^2 (\pi\tl)\, ,
\ee
around the points $x=(2n+1) \pi/\sqrt 2$.

Let us now consider switching on an electric field $e$ along the 
$x^p\equiv x$
direction. In this case $T\propto \tl \cos (x/\sqrt 2)$ is no longer 
a marginal 
deformation, but $T\propto \tl \cos (\sqrt{1-e^2} \, x/\sqrt 2)$ 
is\cite{0208142}. 
The energy-momentum tensor associated 
with the deformed BCFT can be read out from the results of 
\cite{0208142,0301049} by a double Wick rotation, and is given by:
\ben \label{ebc5}
T_{00} &=& {1\over 2} \wt \TT_p \left[ e^2 (1 - e^2)^{-1/2} (1 + 
\cos(2\pi\tl)) + 2 (1-e^2)^{1/2} f(\sqrt{1-e^2}\, x) \right]\, , 
\nonumber 
\\
T_{pp} &=& -{1\over 2} \wt \TT_p \, (1 - e^2)^{-1/2} \, \left( 1 + 
\cos(2\pi \tl)\right)\, ,
\nonumber \\
T_{ij} &=& -(1-e^2)^{1/2} f(\sqrt{1-e^2}\, x) \, \, \delta_{ij} 
\quad
\hbox{for}
\quad 1\le i,j\le (p-1)\, , \nonumber \\
\Pi^p &=& {1\over 2} \wt \TT_p \, e \, (1 - e^2)^{-1/2}(1 +
\cos(2\pi\tl)) \, .
\een
Here $\Pi^p$ denotes the electric flux along the $x^p$ 
direction, or equivalently, fundamental string charge.

If we keep $e$ fixed and take $\tl\to 1/2$ limit, then, as in 
the previous case, $T_{00}$ acquires delta function contribution at 
the points $(2n+1) \pi /\sqrt{2(1-e^2)}$. However, in this limit 
$\Pi^p$ vanishes; thus there is no fundamental string charge left. 
If we want to try to construct a configuration where there is a 
non-zero electric flux along the $x^p$ direction, and at the same 
time take the $\tl\to 1/2$ limit, we must take $e\to 1$ limit 
simultaneously, holding fixed the combination:
\be \label{ebc6}
\Pi^p \simeq {1\over 2} \wt \TT_p \, (1 - e^2)^{-1/2}(1 +
\cos(2\pi\tl)) = \wt \TT_p \, (1 - e^2)^{-1/2} 
\cos^2(\pi\tl)\, .
\ee
Analysing \refb{ebc5} we see that 
in this limit the first term in the expression for $T_{00}$ is equal 
to $|\Pi^p|$, and represents the contribution coming from the 
electric flux, whereas the second term, involving the function 
$f(\sqrt{1-e^2} x)$, goes as
\be \label{ecomp2}
T^{excess}_{00} \equiv T_{00} - |\Pi_p| = {2|\Pi^p| \over (\Pi^p / 
\wt \TT_p)^2 + 2 \tx^2}\, , \qquad \tx = x - {\pi\over 
\sqrt{2(1-e^2)}}\, ,
\ee
for finite $\tx$.
Thus the excess energy density over and above that coming from the 
tension of the fundamental string is no longer strictly localized at 
$\tx=0$.
In particular, for large $\tx$, $T^{excess}_{00}$ falls off as 
$|\Pi^p| 
/ \tx^2$. 
This is perfectly consistent with the results
of section 
\ref{s3} and comparing \refb{ecomp1} with \refb{ecomp2} we see that 
the BCFT results are consistent with a logarithmic growth of the 
tachyon at large $\tx$. This can be taken to be another piece of 
evidence that the 
tachyon effective action given in \refb{ez1} - \refb{evt} correctly 
reproduces the properties of tree level open string 
theory.\footnote{Note that if we chose to examine $T_{\mu\nu}$ and 
$\Pi^p$ for finite $x$ instead of finite $\tx$, we shall get pure 
electric flux tube solutions of \cite{0208142}.}

In fact, if we work with the potential \refb{evt}, then we can 
explicitly reproduce this logarithmic growth in the effective field 
theory using the explicit solutions constructed in \cite{0304180}. We 
can obtain these solutions from those in \cite{0303139} (appendix 
B) by
scaling the $x$ coordinate by $\sqrt{1-e^2}$\cite{0208142}. This 
gives a periodic array of 
kink-antikink solution in the presence of an electric 
field\cite{0304180}:
\be \label{ebcc2}
T = \sqrt 2 \sinh^{-1} \left({a} \, \sin\left(\sqrt{1-e^2} 
{\tx\over 
\sqrt 2}\right)\right)\, , \qquad \qquad \tx = x - {\pi\over
\sqrt{2(1-e^2)}}\, ,
\ee
where $a$ is a parameter labelling the solution. The 
associated value of $\Pi^p$, computed using the method of 
\cite{0208142} on the results of \cite{0303139}, is given 
by\cite{0304180}:
\be \label{ebcc3}
\Pi^p = \wt \TT_p  e (1-e^2)^{-1/2} (1+a^2)^{-1/2}\, .
\ee
The limit we want to consider now is $e\to 1$, $a\to\infty$, keeping 
fixed
\be \label{ebcc4}
\Pi^p \simeq \wt \TT_p (1-e^2)^{-1/2} a^{-1}\, .
\ee
\refb{ebcc2} now gives, in this limit\cite{0304180},
\be \label{ebcc5}
T= \sqrt 2 \, \sinh^{-1} \left( {1\over \sqrt 2} 
{\wt\TT_p \over |\Pi^p|} 
\tx\right)\, .
\ee
This gives the correct logarithmic growth of $T$ at large $\tx$.
The associated value of $T^{excess}_{00}$, computed using the 
effective field theory, is given by\cite{0304180}:
\be \label{ebcc6}
T^{excess}_{00} \equiv T_{00} - |\Pi_p| = {2|\Pi^p| \over 
2(\Pi^p / 
\wt \TT_p)^2 + \tx^2}\, ,
\ee
in qualitative agreement with the exact result \refb{ecomp2}.
The mismatch between \refb{ecomp2} and \refb{ebcc6} of course is the 
result of the well-known mismatch between the results of BCFT and 
effective field theory\cite{0209122,0304045,0304108}. Note however 
that both \refb{ecomp2} and \refb{ebcc6}, after integration over 
$\tx$, 
reproduces the tension of the D-$(p-1)$-brane. This is a reflection 
of the fact that both, the conformal field theory, and the tachyon 
effective action with potential \refb{evt}, correctly reproduces the 
tension of the D-$(p-1)$-brane as a kink solution.

\sectiono{Solution Representing the Fundamental String} \label{s4}

The analysis of section \ref{s3} shows that for a flux tube ending on 
a kink, the tachyon along the flux tube cannot increase faster than 
$x^{1/2}$, $x$ being 
the distance away from the kink. The inability of 
the tachyon to reach its vacuum value $\infty$ within a finite 
distance affects one important property, -- that of exchange 
interactions of the type expected of a fundamental string. Consider, 
for example, two segments of flux tubes, $APB$ and $DPC$, 
intersecting at a point $P$. Let us further suppose that $APB$ 
represents a segment of a flux tube at a distance $d_1$ from a kink 
on which it ends, and $DPC$ represents a segment of a flux tube at a 
distance $d_2$ from a different kink on which it ends.  If $d_1$ and 
$d_2$ are both large but different, then the values of $T$ inside the 
segments $APB$ and $DPC$ will also be large but different. Let us 
denote them by $T_1$ and $T_2$ respectively. 

Now consider the exchange process by which the system described above 
gets converted to two new segments $APC$ and $DPB$. Fundamental 
strings are allowed to have such exchanges. However in this case, 
inside $APC$ and $DPB$, the tachyon must jump from $T_1$ to $T_2$ (or 
$T_2$ to $T_1$) across the point $P$. Since this costs energy, such 
processes will not be energetically favourable. Thus we see that 
even if we are able to construct electric flux tube solutions for 
which $T$ grows as a power law along the flux tube, such solutions 
will be missing one important property of the fundamental string.

Using the insight gained from the analysis of section \ref{s3}, we 
shall now explore the possibility of constructing a 
different type of string-like 
solution which carries
electric 
flux as is required of a fundamental string, can end on the kink, 
and can also have exchange interactions of the kind described above.
In order that the string can end on a kink, the flux lines 
inside the string must be able to smoothly match the 
electric flux 
lines inside a kink which flow radially 
outwards from the point where the 
string ends. 
Since the lower bound to the energy given in \refb{ewidth3} is 
proportional to the component of $\vec\nabla T$ along the flux line,
we can try to avoid this energy cost by making the
flux lines follow
constant tachyon profile.
If we follow this approach, then 
the $\vec\Pi$ 
across a 
cross section of the fundamental string should be correlated with 
$T$ in the same manner in which $\vec\Pi$ and 
$T$ are correlated inside a kink via
eq.\refb{ewidth2c}. 
Since inside a D-brane most 
of the contribution to the electric flux comes from regions where $T$ is 
finite, the same must be true for the fundamental string configurations. 
In other words, in order to embed a fundamental 
string solution in the 
tachyon vacuum, we should create a region of finite $T$ and embed 
the 
electric flux in this region. Since it costs energy to create a region of 
finite $T$ due to non-zero value of the tachyon potential and the 
derivative of the tachyon, we must 
minimize the energy. This requires the volume of this region to vanish. 
Thus unlike the flux tube solution of \refb{ewidth2a}, these new 
configurations 
cannot spread over a finite $(p-1)$-volume transverse to the direction of 
the flux.

In fact an explicit construction of such a configuration which can end 
on the 
kink is already available. These are the solutions given in 
\cite{9708147,9709027,9709014,9711094}. In these papers it was shown 
that in the 
presence of 
a point electric charge source on a D-$(p-1)$-brane, the DBI action 
on the  
brane admits a solution which has the interpretation that the brane 
gets
deformed into the form of a long hollow tube attached to the original 
brane like a 
spike, with 
the 
electric flux flowing along the wall of the tube.
In the world-volume theory on the D-$(p-1)$-brane the solution of 
ref.\cite{9708147} takes 
the form:
\be \label{ecmsol}
\pi^s = \pm A \, \TT_{p-1} \, \, {\xi^s \over r^{p-1}}, \qquad y^p = 
{A\over p-3} \, \, {1 \over
r^{p-3}}\, , \qquad r = \sqrt{\sum_{s=1}^{p-1} \xi^s \xi^s}\, ,
\ee
where $A$ is a 
constant labelling the total amount of flux carried by the solution. 
The gauge field $a_s(\xi)$ associated with this solution is 
determined from its equation of motion.
{}From this we see that as $r\to 0$, $y^p\to\infty$, {\it i.e.} we 
move further and further away from the plane of the D-$(p-1)$ brane 
($y^p=0$). For small $r$
the D-$(p-1)$-brane  
looks like $\RR\times S^{p-2}$, and
the radius of $S^{p-2}$ decreases as we go away from the plane of the 
original D-$(p-1)$-brane. When the constant $A$ is adjusted so that 
the electric flux takes its minimum value consistent with the 
quantization laws, the solution
was given the interpretation of a fundamental string ending on a 
D-$(p-1)$-brane, and its 
world-volume dynamics, quantum numbers and tension were shown to 
be consistent with this interpretation.
Since all 
solutions of the DBI action on a BPS D-$(p-1)$-brane can be lifted to a 
solution of the tachyon effective action on a non-BPS 
D-$p$-brane\cite{0303057}, we can 
now translate the solution of the DBI theory into a solution in the 
tachyon effective field theory. 
Using \refb{eprofile}, \refb{ewidth2c}, and \refb{ecmsol} we get,
\ben \label{edd1}
T &=& a F \left(x^p - {A\over p-3} \, \, {1\over 
r^{p-3}}\right)
\, , 
\nonumber \\
\Pi^s&=& \pm A \, {\xi^s\over r^{p-1}} \, a F' \left(x^p - 
{A\over p-3} \, \, {1\over r^{p-3}}\right) \, V\left(a F\left(x^p -
{A\over p-3} \, \, {1\over r^{p-3}}\right) \right)\, , 
\nonumber \\
\Pi^p &=& \pm A^2 \, {1\over r^{2p-4}} \, a F' \left(x^p -
{A\over p-3} \, \, {1\over r^{p-3}}\right) \, V\left(a F\left(x^p -
{A\over p-3} \, \, {1\over r^{p-3}}\right) \right)\, , 
\nonumber \\
&& (\xi^0, \ldots \xi^{p-1}) \equiv (x^0, \ldots x^{p-1}), \qquad r = 
\sqrt{\sum_{s=0}^{p-1} \xi^s \xi^s} = \sqrt{\sum_{s=0}^{p-1} x^s 
x^s}\, , 
\een
and $A_\mu$ is any smooth vector field whose pullback on the surface 
$x^p = A / \{ (p-3) \, r^{p-3} \}$ is equal to $a_s$ computed from 
\refb{ecmsol}.\footnote{Note that since there is no magnetic flux on 
the kink world-volume, there is no topological obstruction to 
choosing such a smooth vector field. We could simply choose an $A_\mu$ 
which has the required value on the world-volume of the kink, 
and goes to zero quickly as we move away from the location of 
the kink.} 
In this solution the fundamental string 
corresponds to a tubular region with the wall of the tube having a 
thickness of order $1/a$. Of course we need to take the $a\to\infty$ 
limit in order to ensure that \refb{edd1} gives a solution of the 
equations of motion derived from the Hamiltonian \refb{ewidth1}. 
Analogous solutions in boundary string field theory have been considered 
earlier in ref.\cite{0102174}.

This provides a description of the fundamental open string that can 
end on
a kink. As we go away from the plane of the kink, the radius of the 
tube 
decreases. Thus the description of a 
fundamental string far away from the 
kink is given as a rolled up kink solution with infinitesimal radius 
$R$.
However, in describing the solution we need to take the $a\to\infty$, 
$b\to 0$ and $R\to 0$ limit in this specific order.\footnote{These 
solutions
may also be considered as the zero magnetic field limit of 
supertubes\cite{0103030}. However, supertubes themselves, carrying 
electric and magnetic flux, cannot be considered as non-singular 
solutions in the 
tachyon effective field theory, since due to the presence of the magnetic 
flux on the kink world-volume there is now a topological obstruction to 
continuing the gauge field smoothly in the region inside the tube.} 
In order to see how the energetics work out, we can use the language of 
the world-volume theory of the D-$(p-1)$-brane. A straight fundamental 
string will then be described by a D-$(p-1)$-brane world-volume of the 
form $\RR\times S^{p-2}$, with electric field along the direction of 
$\RR$. 
If 
we assume for simplicity that the electric field is 
uniform, and has magnitude $e$, and if $\VV$ denotes the 
$(p-2)$-volume of 
$S^{p-2}$, then the total flux is given by $\TT_{p-1}\VV e / 
\sqrt{1-e^2}$ and must 
be fixed if we are to describe a given number (say 1) of fundamental 
strings.
On the other hand the energy per unit length 
along $\RR$ for this configuration is given by:
\be \label{etoten}
\EE = \TT_{p-1} \VV / \sqrt{1 - e^2} \, .
\ee
Thus in order to get a minimum energy configuration for a given 
fundamental string charge, we need to minimize $\VV/\sqrt{1-e^2}$ 
keeping 
$\VV e/\sqrt{1-e^2}$ fixed. This leads to the limit,
\be \label{elimit}
\VV\to 0, \qquad e\to 1, \qquad \VV/\sqrt{1-e^2} = \hbox{fixed}\, .
\ee

This 
description makes it clear that (within this approximation) as long 
as we 
take the D-$(p-1)$-brane to roll up in the configuration $\RR\times 
S^{p-2}$, 
the 
fundamental open string has zero width, since requiring the volume of 
$S^{p-2}$ to be 0 implies that its radius must go to zero. 
It is also easy to see following \cite{0009061,0010240} that the 
world-volume dynamics of such a string is 
described by the Nambu-Goto action. For this we need to recall that the 
analysis of \cite{0009061,0010240} was carried out under the 
assumption 
that the $V^2\det h$ term in \refb{ewidth1} can be ignored in the 
study of 
the 
dynamics of the flux tube, and this was sufficient to establish that 
the
world-volume dynamics of an infinitely thin flux tube is governed by 
the 
Nambu-Goto action without any higher derivative corrections. 
Thus all we need to show is that the $V^2\det h$ term in \refb{ewidth1} 
can be 
neglected for studying the dynamics of flux tubes associated with rolled 
up kink solution. Now, from \refb{ewidth2c} we see that the $\Pi^s\Pi^s$ 
term 
in \refb{ewidth1} is of order $(\TT_{p-1})^{-2}\, (aF' V)^2 \pi^s 
\pi^s$.
On the other hand, the $V^2\det h$ term in \refb{ewidth1}, which is 
dominated by the $\p_i T \p_j T$ 
term in $h_{ij}$, goes as $V^2 (a F')^2$. Thus in the limit 
\refb{elimit}, 
in which $\pi^s \sim e / \sqrt{1-e^2}$ blows up, we have
\be \label{egreater}
\Pi^s \Pi^s >> V^2 \det h\, .
\ee
Thus we can ignore the $V^2 \det h$ term in the analysis of the dynamics 
of the fundamental string. This, in turn, establishes that the 
dynamics is 
governed by the Nambu-Goto action.

The above construction provides a description of the open string 
ending on a kink. Given this description of the fundamental open 
string solution, 
fundamental closed 
strings 
can be described as closed loops of such rolled up kink solutions. 
When two such (open or closed) strings cross they can have the usual 
interaction in which two segments $APB$ and $DPC$ of fundamental string, 
crossing at a point $P$, can become another pair of segments $APC$ and 
$DPB$. 
This is possible because the profile of the tachyon across a 
cross-section of the tubes representing $APB$ and $DPC$ are 
identical, except possibly a small difference in the radii of the 
tubes if they represent segments of open strings which are at 
different distances from the kinks on which they end.
In contrast if a segment $APB$ of an electric flux tube in whose 
core $T=\infty$ crosses a segment $DPC$ 
of a rolled up kink solution then 
they cannot have this type of exchange interaction, since the would be 
final configurations $APC$ and $DPB$ will involve electric flux 
travelling 
from the $T=\infty$ vacuum to finite $T$ region and is energetically 
unfavourable. This shows that it is inconsistent to identify the closed 
strings as the electric flux tube solutions with $T=\infty$ core if we 
have identified the open strings 
as rolled up kink solutions.

Before concluding this section we would like to add a word of
caution. In describing the fundamental string as a rolled up kink
solution, we could in principle replace $S^{p-2}$ by another compact
$(p-2)$-dimensional space $K_{p-2}$ whose volume vanishes, but which
nevertheless has some dimensions finite. An example of such a compact
space could be simply an elongated sphere (ellipsoid) of the form
$R^2 (x^1)^2 + (x^2)^2 + \ldots (x^{p-1})^2 = R^2$, and we take the
$R\to 0$ limit. The energetic considerations do not prevent us from
having such a configuration, and this will describe a configuration
in which the charge of a fundamental string along $x^p$ spreads over
the $x^1$ direction. This clearly violates the known property of the
fundamental string. We believe the resolution of this puzzle must
come from taking into account higher derivative corrections / quantum
corrections to the action \refb{ez1}, since the situation that we are
describing now is simply that of a rolled up BPS D-$(p-1)$-brane with
electric flux, and there must be an underlying mechanism that
prevents the brane to collapse in a manner that allows us to spread
out a single unit of electric flux over a subspace of dimension $>1$.
To this end note that the configuration that we are considering is
far outside the domain of validity of the effective action
\refb{ez1}, and only the BPS nature of the configuration can
guarantee that the solution survives higher derivative / quantum
correction. Requiring the configuration to be BPS at the quantum
level could certainly fix the shape of the transverse section.

\sectiono{Closed Strings and Decaying D-branes} \label{s5}

Given that the electric flux tube solution given in 
\refb{ewidth2a} cannot be used to describe a single 
fundamental string, one could ask what could be the possible physical 
interpretation of these solutions. In this section we shall propose a 
possible answer to this question. However we begin our discussion by 
trying to find the physical interpretation of another related 
system, -- 
the tachyon matter produced by a rolling tachyon at late time
\cite{0203211,0203265,0204143}. 
Both the 
electric flux tube solution and the tachyon matter are characterized 
by 
the fact that they involve configuration where the tachyon remains large 
(near its vacuum) everywhere in space.
 
The rolling tachyon solution describes the process of classical decay of 
a brane-antibrane system or a non-BPS D-brane.
Classical analysis indicates that the rolling of the tachyon 
on these systems produces at late time
a state 
of non-zero
energy density, concentrated on the plane of the original brane, and
vanishing pressure. 
In particular for the decay of a non-BPS D$p$-brane, the energy density 
$\EE$ and the pressure $p_\parallel$ and $p_\perp$ along directions 
tangential and transverse to the brane respectively have 
the form:
\be \label{eco1}
\EE = {C\over g}\delta(\vec x_\perp), \qquad p_\parallel(x^0) = 
{1\over g} 
f(x^0,C) \delta(\vec x_\perp)\, , \qquad p_\perp(x^0) =0\, .
\ee
where $C$ is some constant labelling the initial condition on the 
tachyon field, $g$ is the string coupling, $\vec 
x_\perp$ denote directions transverse to the D$p$-brane 
world-volume, and 
$f(x^0,C)$ is a function computed in \cite{0203211,0203265} which 
vanish 
for large $x^0$. 

The natural question to ask now is: what is the physical
interpretation of this system? Naively, since there are no physical
open string states around the tachyon vacuum, one would expect that
the D-brane should decay into a collection of closed strings. On the
other hand since closed strings appear in the open string loop
expansion\cite{thorn}, one would expect that the effect of emission
of closed strings should already be contained in the quantum open
string theory, and one should not have to include the effect of
closed string emission as an additional contribution beyond what
quantum open string theory gives us. Keeping this in view let us now
try to see how quantum open string theory will modify the classical
results \refb{eco1}, and then try to compare these with the expected
answer that we get assuming that the unstable D-brane system decays
to closed strings.

According to Ehrenfest theorem, the classical
evolution of the energy momentum tensor given in \refb{eco1} should 
reflect the
evolution of the expectation value of the energy momentum tensor
in the zero coupling limit. 
In other words, during the decay of a
D-brane the quantum expectation values of energy density and pressure
should follow this classical answer in the weak coupling limit. 
However,
for a finite coupling quantum effects will modify these classical 
results 
by modifying 
the effective action, and hence modifying the effective equations of 
motion. From general 
considerations we should then expect the quantum corrected results 
for the 
evolution of $\EE$, $p_\parallel$ and $p_\perp$ to be of the form:
\be \label{eco2}
\EE = {1\over g} \ve(\vec x_\perp, x^0, g, C)\, ,
\qquad p_\parallel(x^0) = {1\over g} \phi_\parallel(\vec x_\perp, x^0, 
g,C)\, ,
\qquad p_\perp(x^0) = {1\over g} \phi_\perp(\vec x_\perp, x^0, 
g, C)\, ,
\ee
where $\ve$ and $\phi$ are functions which are in principle computable 
by 
quantizing the theory in the rolling tachyon background. 
The precise form of the functions $\ve$ and $\phi$ may depend on the 
choice of the `vacuum state' used for this computation since there 
is no 
natural choice of the vacuum state for time dependent background, and we 
have many different quantum states corresponding to the same classical 
configuration.
However, 
in 
the $g\to 0$ limit, we must have
\be \label{eco3}
\ve(\vec x_\perp, x^0, g, C)\to C \, \delta (\vec x_\perp)\, , \qquad
\phi_\parallel(\vec x_\perp, x^0, g, C)\to f(x^0)\, \delta (\vec 
x_\perp)\, ,
\qquad \phi_\perp(\vec x_\perp, x^0, g, C)\to 0\, .
\ee

Let us now compute the expected contribution to the energy momentum 
tensor 
associated with the closed strings to which the D-brane should decay 
and 
compare this with \refb{eco2}, \refb{eco3}.\footnote{Of course, since 
the closed string sector includes gravity, the definition of 
stress tensor has the usual problem. However, since only a very small 
fraction of the energy goes into graviton states, we could consider 
the contribution to the energy-momentum tensor from the 
non-gravitational sector of the  closed string.}  
The computation of closed string emission from unstable D-branes
has been carried out in detail recently\cite{0303139,0304192} where 
it was found that  
for D-$p$-branes for 
$p\le 2$, the emission of closed 
strings from this background extracts all the energy of the original 
brane 
into closed string modes.\footnote{This analysis only deals with the 
closed string states created from the vacuum by space-like 
oscillators. There may also be interesting information in the closed 
string states associated with time-like oscillators\cite{0208196}, 
but we 
shall ignore them in the present discussion.}
In particular,
the final state for the decay of a non-BPS D0-branes is
dominated by highly non-relativistic closed strings of mass $\sim 
g^{-1}$ 
and velocity of order $g^{1/2}$. Although for non-BPS D-$p$-branes for 
$p\ge 2$ naive analysis involving homogeneous rolling tachyon tells us 
that only a small fraction (of order $g$) of the D-brane energy is 
radiated away into closed strings, it was argued in \cite{0303139} 
that in 
the 
presence of any
inhomogeneity, the decay of any D-$p$-brane for $p\ge 1$ can be 
thought of 
as the 
result of decay of a collection of
non-BPS D0-branes. 
Hence its final state will also be 
dominated by highly non-relativistic closed strings of mass $\sim 
g^{-1}$ 
and velocity of order $g^{1/2}$.
Even for spatially homogeneous configuration, it was shown in 
\cite{0304192} that higher moments of the energy density of the final 
state closed 
strings diverge in the $g\to 0$ limit, although the mean value
is finite. Thus there is a large uncertainty in the energy of the 
emitted closed strings, and presumably, once quantum corrections are 
included, one would find that even a homogeneous configuration decays 
to closed strings of energy of order $g^{-1}$.\footnote{I would like 
to thank L.~Rastelli for discussion on this point.}

Since
the closed strings 
produced during the decay of the non-BPS D$p$-brane have velocity of 
order 
$g^{1/2}$, it takes a  time of order $g^{-1/2}$ for these closed 
strings to carry the energy away from the original location of the 
brane. 
This is perfectly consistent with \refb{eco2}, \refb{eco3}. In 
particular 
a specific choice of $\ve(\vec x_\perp, x^0, g, C)$ which satisfies 
\refb{eco3} is $C\left({\pi\over g (x^0)^2}\right)^{n_\perp/2} 
\exp\left(-{\vec x_\perp^2 \over g 
(x^0)^2}\right)$ where 
$n_\perp$ is the number of transverse dimensions. In this case for any 
finite $x^0$ as we take the $g\to 0$ limit we shall see the energy 
density 
localized on the plane of the original brane, whereas over a period of 
order $g^{-1/2}$ it disperses to a distance of order one in the 
transverse 
directions.

In fact, the actual rate of dispersal of the energy away from the 
plane of the brane may be even slower due to the gravitational 
attraction that tends to pull the decay products towards the plane of 
the brane. Since the Newton's constant is 
of order $g^2$, for two objects of mass 
$\sim g^{-1}$ separated by a distance of order 1, the escape velocity 
is of order $g^{1/2}$. However, since in the present
case the decay products are initially localized within a 
smaller distance from the plane of the brane, the escape velocity 
will be larger than $g^{1/2}$. Thus the decay products, with a 
typical velocity of order $g^{1/2}$, will not be able to escape to 
infinity and will be pulled back towards the plane of the brane.

This argument shows that the massive decay products by 
themselves cannot carry the energy away from the plane of the brane. 
But one might expect
that these very massive closed strings will 
eventually decay into massless states which carry the energy away 
from the plane of the brane. Note however that due to the 
exponentially growing density 
of states at high mass level, 
a very massive string state will decay 
predominantly to other very massive string states unless such decay 
processes are suppressed by exponentially small matrix elements. Due 
to the presence of a large number of 
such massive closed strings near the plane of the original brane, 
these 
strings will collide frequently, producing 
(predominantly) massive closed string states. Thus the tachyon 
matter, describing the decay 
product of a non-BPS D-brane, may be longer lived than one would 
naively expect it to be. (Incidentally, it will be interesting to 
see if this argument can be sharpened to estimate the life-time of 
massive 
black holes represented by elementary string 
states\cite{9309145,9401070,9405117,9504147,9506200,9612146}.)

Let us now turn to the analysis of the pressure. For a collection of
non-relativistic particles, the ratio of the pressure to the energy
density is proportional to the square of the average velocity of the
particles. As mentioned in the previous paragraph, for the closed
strings produced in the decay of non-BPS D-branes this is of order
$g$.  Thus after all the energy of the D0-brane has been converted to
the closed string states, the pressure of the system will be of order
unity, since the energy density is of order $g^{-1}$.  This agrees
perfectly with the result of \refb{eco2}, \refb{eco3} which states
that asymptotically, the order $1/g$ contribution to the pressure
vanishes since $f(x^0)\to 0$ as $x^0\to\infty$.

This suggests that the classical tachyon matter, produced during the
decay of an unstable D-brane, may be the open string description of a
collection of highly non-relativistic closed strings of high density
that is expected to be produced in this decay.  Given that closed
strings appear at open string loop level, it may seem somewhat
surprising that tree level open string theory contains information
about properties of closed strings.  However this could be a
reflection of the fact that since there are no open string states
around the tachyon vacuum, even the classical open string theory must
know something about closed strings whose average property it is
supposed to reproduce in the weak coupling limit\cite{0304192}. This
interpretation is consistent with the idea that quantization of open
string theory around the tachyon vacuum should give rise to closed
string theory\cite{9904207,0111092,0111129,0207266,0304224}.

One of the lessons we can learn from this interpretation is that the
classical results are quite unreliable when the energy density of the
system falls below the string density. In particular classical
analysis tells us that the ratio of pressure to the energy density
vanishes even at energy density below the string scale, but this is
not expected to happen in the quantum theory since below string
density the system should behave as ordinary radiation.  {}From the
general form \refb{eco1} we see that this happens because in this
case the quantum correction, which are of order 1, could dominate the
classical contributions to $T_{\mu\nu}$.

Let us now turn to the interpretation of the electric flux tube
solution of section \ref{s2}. For convenience let us compactify the
direction along which the electric flux points, so that
the fundamental string charge associated with the electric flux has a
simple interpretation of fundamental string winding number. In
analogy with the rolling tachyon solution, we should expect that the
classical results will be a good approximation to the complete
answers when the string coupling $g$ is small, and the energy density
and the winding number density is of order $g^{-1}$. This is large 
for small $g$. In this case we
could interprete the freedom of spreading out the winding number
simply to the possibility of distributing these large number of
fundamental strings arbitrarily in the hyper-plane transverse to the
compact direction. The classical dynamics of the tachyon effective
field theory then describes the time evolution of the expectation
values of various physical quantities for such a system.

The analysis of section \ref{s3a} shows that in the presence of such
electric flux, a D-$(p-1)$-brane placed transverse to the flux will
fatten with a width of order $|\vec\Pi| / \wt\TT_p$. We could ask if
there is a physical understanding of this fattening based on our
interpretation given above. It is tempting to suggest that this
fattening is caused by a large number of fundamental strings ending
on the D-$(p-1)$-brane from both sides. Since fundamental string
deforms the D-brane into the shape of a spike as described in 
section \ref{s4}, the average effect of
a large density of spikes on both sides of the D-brane will be to
effectively fatten the D-brane. Since the total number of spikes is
proportional to the total flux $|\vec\Pi|$, this will easily explain
why the excess energy density away from the original plane of the
D-$(p-1)$-brane is proportional to $|\vec \Pi|$.

\sectiono{Summary} \label{s6}

We conclude the paper by summarizing the main results.

\begin{enumerate}

\item The tachyon effective field theory is 
known to contain electric flux 
tube solutions for which the electric field is at its critical value 
and
the tachyon is at infinity. These flux 
tubes have many properties in common with the fundamental strings. We 
show that if such a  
flux tube `ends' on a kink solution of the 
effective field theory representing a BPS D-$(p-1)$-brane, then the 
tachyon cannot increase faster than $x^{1/2}$ as we move a distance 
$x$ away from the plane of the kink along the flux tube. Thus the 
flux tube approaches its asymptotic configuration, where the tachyon 
is at its vacuum value $\infty$, very slowly. \label{ii1}
There is no explicit classical solution known at present 
describing 
such configurations, but we have argued that even if we are able to 
construct  such solutions, they will be missing out on one important 
property of the fundamental string, -- that of exchange interaction.

\item We propose an alternative form of the electric flux tube
solution for which the tachyon is finite in the region which carries
the electric flux.  Energetic consideration then forces this region
to have zero volume.  While this is an improvement over the electric
flux tube solutions mentioned in item \ref{ii1}, this constraint is
not sufficiently strong to confine the electric flux to a one
dimensional subspace, as is required if it has to describe a single
fundamental string. We suggest that higher derivative / quantum
corrections, which are expected to be significant for these
solutions, could be responsible for localizing the fundamental string
charge to a one dimensional subspace.

\item A recent analysis of time dependent solutions describing the
decay of a non-BPS D-brane has suggested that during the decay
process all the energy of the D-brane is converted to closed strings.
We suggest that this is an alternative description of the phenomena
that we see in the analysis in open string theory, and verify this by
comparing the properties of the tachyon matter obtained in the
classical open string analysis with the properties of the system of
closed strings expected to be produced during the decay process. This
analysis suggests that tachyon matter in effect describes a system of
closed strings at high energy density. In the same spirit, we also
suggest that the electric flux tube solutions described in item
\ref{ii1} represent a system of closed strings at high energy density
and high winding charge density. \label{ii3}

\end{enumerate}

{\bf Acknowledgement}: I would like to thank K.~Hashimoto, Y.~Kim, 
L.~Rastelli and S.-J. Rey
for useful discussions.

\end{document}